# Structural identifiability of compartmental models for infectious disease transmission is influenced by data type

Author names and affiliations: Emmanuelle A. Dankwa[1], Andrew F. Brouwer[2], Christl A. Donnelly[1,3]

[1] Department of Statistics, University of Oxford, Oxford, United Kingdom
[2] Department of Epidemiology, University of Michigan, 1415 Washington Heights, Ann Arbor, MI 48109-2029, USA
[3] Department of Infectious Disease Epidemiology, Faculty of Medicine, School of Public Health, Imperial College London, United Kingdom

Corresponding author: Emmanuelle A. Dankwa, Department of Statistics, University of Oxford, 24-29 St Giles', Oxford OX1 3LB, United Kingdom

Email address: dankwa@stats.ox.ac.uk

**ABSTRACT**

If model identifiability is not confirmed, inferences from infectious disease transmission models may not be reliable, so they might lead to misleading recommendations. Structural identifiability analysis characterizes whether it is possible to obtain unique solutions for all unknown model parameters, given the model structure.

In this work, we studied the structural identifiability of some typical deterministic compartmental models for infectious disease transmission, focusing on the influence of the data type considered as model output on the identifiability of unknown model parameters, including initial conditions.

We defined 26 model versions, each having a unique combination of underlying compartmental structure and data type(s) considered as model output(s). Four compartmental model structures and three common data types in disease surveillance (incidence, prevalence and detected vector counts) were studied.

The structural identifiability of some parameters varied depending on the type of model output. In general, models with multiple data types as outputs had more structurally identifiable parameters, than did models with a single data type as output.

This study highlights the importance of a careful consideration of data types as an integral part of the inference process with compartmental infectious disease transmission models.

## 1.1 INTRODUCTION

Goals of infectious disease transmission modelling often include making inferences about the underlying transmission process, predicting the future course of an epidemic given a range of interventions, or estimating what would have happened in a counterfactual scenario. A defined model is fitted to a given data set (frequently an incidence time series generated by passive surveillance). This model-fitting process is parameter estimation, where one determines parameter values or distributions corresponding to model outputs that best fit (or at least, approximate) the observed data. Parameter estimation however can only produce robust results if the model is *identifiable* (Audoly et al., 2001; Castro and de Boer, 2020; Cobelli and Distefano III, 1980; Kao and Eisenberg, 2018; Ljung and Glad, 1994; Villaverde et al., 2016; Wieland et al., 2021); that is, if it is possible, in principle, to obtain unique solutions for all unknown model parameters, given the model structure and available data. We note that other properties such as predictability (Castro et al., 2020; Scarpino and Petri, 2019) and uncertainty quantification (Capaldi et al., 2012; McCabe et al., 2021) also affect the reliability of model inferences (Massonis et al., 2021a); these are not treated here, however.

Although the subject of identifiability has received considerable attention in the systems biology and control literature (see (Wieland et al., 2021) for a recent review), it is inconsistently applied in the infectious disease modelling literature. Relatively few studies exist on the identifiability analysis of infectious disease models (Brouwer et al., 2018; Eisenberg et al., 2013; Evans et al., 2005; Kao and Eisenberg, 2018; Massonis et al., 2021a; Tuncer et al., 2016; Tuncer and Le, 2018) and the practice of routinely checking the identifiability of these models *before* parameter estimation is not widespread. Nevertheless, identifiability is required to make meaningful inferences on model parameters and, consequently, to provide reliable evidence to inform public health policymaking.

In a non-identifiable model, parameter sets with similar values may yield considerably different model predictions (Kao and Eisenberg, 2018; Roda et al., 2020). Thus, a failure to consider identifiability could result in misleading recommendations, as has been previously noted (Kao and Eisenberg, 2018; Massonis et al., 2021; Roda et al., 2020; Roosa and Chowell, 2019), some of which could have serious consequences. For example, Kao and Eisenberg demonstrated using a dengue transmission model that two sets of parameters which fit the incidence data comparably well yielded very different predictions for incidence after an intervention is applied (see Fig. 9 in their paper) (Kao and Eisenberg, 2018). Roda and colleagues also showed that the lack of identifiability in COVID-19 transmission models could lead to extreme variability in predictions (Roda et al., 2020).

A distinction is made between the two types of identifiability: *structural identifiability* and *practical identifiability*. Structural identifiability, a concept first introduced by Bellman and Astrom in 1970, is a property of the model structure and associated measurement function (i.e., the function of model variables that is to be observed) and does not depend on the quantity or quality of the observed data. It addresses the question: Given an error-free model structure, and assuming noise-free, infinite data, do unique solutions exist for the model parameters? Structural identifiability is affected by: 1) the nature of the model parameterization (Muñoz-Tamayo et al., 2018) which influences symmetries, i.e., functional relationships between model parameters (Eisenberg and Hayashi, 2014; Hengl et al., 2007; Massonis et al., 2021a; Villaverde, 2022); and 2) the data type considered as model output (Balsa-Canto et al., 2010; Chis et al., 2011; Massonis et al., 2021a; Tuncer and Le, 2018). Practical identifiability, on the other hand, is related to the adequacy of the available observed data for the estimation problem at hand (Balsa-Canto et al., 2010; Brouwer et al., 2017; Kao and Eisenberg, 2018; Miao et al., 2011; Raue et al., 2009; Tuncer et al., 2016; Tuncer and Le, 2018). The corresponding question is: Do the data contain enough information to infer the model parameters? Structural identifiability is a necessary, but not sufficient, condition for practical identifiability; that is, a structurally non-identifiable parameter cannot be practically identifiable, and a structurally identifiable parameter could be practically non-identifiable depending on the data available (Cobelli and Distefano III, 1980; Eisenberg et al., 2013). In this work, we are considering structural identifiability of infectious disease transmission models.

Several studies have demonstrated the influence of the type of observed data on the structural identifiability of infectious disease models. For example, Tuncer and Le studied a Susceptible-Infected-Treated-Recovered epidemic model which becomes structurally identifiable only when both cumulative incidence rates and the number of treated individuals is observed (Tuncer and Le, 2018). In the same work, the authors explored the identifiability of a Susceptible-Exposed-Infected-Recovered model and showed that the type of structural identifiability for two parameters (recovery rate and length of latent period) depended on whether the observed data were cumulative incidence or prevalence. Similar works include (Evans et al., 2005), on the structural identifiability of a seasonally forced SIR model with prevalence and a proportion of the incidence as outputs; (Eisenberg et al., 2013), on the identifiability of parameters of compartmental models for cholera with prevalence as output; (Tuncer et al., 2016), on the identifiability of an immune-epidemiological model for Rift Valley fever with time-series data of viremia levels as output; (Kao and Eisenberg, 2018), on the identifiability of a dengue transmission model with various types of human and mosquito incidence data as outputs; and more recently, (Massonis et al., 2021a), on the structural identifiability of a wide range of COVID-19 transmission models with a variety of surveillance data types as outputs.

However, few of these studies (e.g., (Eisenberg et al., 2013; Evans et al., 2005)) have explicitly studied the identifiability of unknown initial conditions (ICs). Other studies have either assumed known ICs (e.g., (Tuncer and Le, 2018)) or have implicitly considered unknown ICs through assessment of the *observability* of model states (Massonis et al., 2021a); i.e., whether the state variable trajectories can be uniquely determined from observed data. (Structural identifiability has been considered as a particular case of observability (Massonis et al., 2021a; Sedoglavic, 2002; Tunali and Tarn, 1987; Villaverde, 2019).) Here, we explicitly consider ICs as unknown parameters in all models and analyse their structural identifiability given various data types. Often values are assumed for ICs, but careful analysis often reveals that parameter estimates depend on these IC assumptions. We can ask under what circumstances ICs can be uniquely determined from observed data. Although this question might technically be considered one about observability, when the ICs are reframed as parameters, the question is one of identifiability. Thus, our work adds to the literature by examining how the structural identifiability of ICs of classic compartmental models change with data type. Additionally, we employ a publicly available web-based toolbox, SIAN (Hong et al., 2019), to analyse the structural identifiability of model parameters, allowing us to demonstrate the utility of such tools. Specifically, we consider four compartmental structures (SIR, SLIR, SLIR with vaccination and relapse and a vector-borne disease model with SLIR for hosts and SLI for vectors) and three common data types in disease surveillance (incidence, prevalence and detected vector counts). Using SIAN, we analyse the structural identifiability of unknown parameters in 26 model versions, each a unique combination of underlying compartmental structure and data type considered as model output. We use the term "model version" to refer to a compartmental structure-output(s) combination; e.g., SIR with incidence, or SLIR with incidence and prevalence.

Although the compartmental structures and data types we consider are by no means exhaustive, our work is intended to demonstrate the importance of identifiability and to be instructive for those seeking to apply these techniques to their own models. We have therefore made available all input codes and output files to facilitate reproducibility: structural-identifiability-epi-models · GitFront.

The chapter is outlined as follows. In Section 1.2, we introduce the general modelling framework and notation and provide formal definitions of relevant structural identifiability concepts. Here, we also introduce the four compartmental structures, briefly introduce the software toolbox utilized, present the model versions examined and finally, outline the structural identifiability analysis performed. Section 1.3 presents the results and Section 1.4 presents a discussion of results. Concluding remarks are given in Section 1.5.

## 1.2 METHODS

### 1.2.1 General modelling framework and formal definitions

Consider a deterministic ordinary differential equation (ODE) infectious disease transmission model $\mathcal{M}$ of the form

$$\mathcal{M} := \begin{cases} \dot{X}(t) = f(X(t), p, u(t)) \\ y(t) = g(X(t), p) \\ X_{t_0} = X(t_0) \end{cases}, \tag{1}$$

with observations on the interval $t_0 \leq t \leq T$, where $\dot{X}(t)$ is a system of non-linear ODEs, $X(t) \in R^{n_X}$ is a vector of time-varying disease states and the unique solution to the system $\dot{X}(t)$, $p \in R^{n_p}$ is a vector of constant unknown model parameters, $y(t) \in R^{n_y}$ is a vector of time-dependent model outputs corresponding to a specific data type (for example, case incidence rates), $u(t) \in R^{n_u}$ is a time-dependent input vector, $g$ is the measurement equation (which defines the relationship between $X(t)$, $p$ and $y$), and $X_{t_0} \subset R^{n_X}$ is a vector of known ICs. Note that unknown components of $X_{t_0}$ are included in $p$ and that $f$ and $g$ are vectors of analytic functions of their arguments. The formal definition of structural identifiability for a model and its parameters is given below. The structural identifiability of a parameter may either be *local* (i.e., holding only within a limited region of the parameter space or about a given point) or *global* (i.e., holding (almost) everywhere within the parameter space) (Ljung and Glad, 1994).

**Definition 1** [Parameter structural identifiability] (Cobelli and Distefano III, 1980; Ljung and Glad, 1994)

*A parameter $p_i \in p$ is <u>structurally globally identifiable (s.g.i.)</u> on $[t_0, T]$ for a given output choice $y$ if a unique solution exists for $p_i$; that is, if and only if for almost any $p^*$ and almost any IC (i.e., excluding degenerate values), $y(X, \hat{p}) = y(X, p^*)$ implies $\hat{p}_i = p_i^*$. Otherwise, $p_i$ is <u>structurally globally non-identifiable</u>.*

*A parameter $p_i \in \boldsymbol{p}$ is <u>structurally locally identifiable (s.l.i.)</u> on $[t_0, T]$ for a given an output $\boldsymbol{y}$ if there exists a neighbourhood $V(\boldsymbol{p})$ of the parameter space within which a unique solution exists for $p_i$. Otherwise, $p_i$ is <u>structurally non-identifiable (s.n.i.)</u>.*

**Definition 2** [ Model structural identifiability] (Cobelli and Distefano III, 1980; Ljung and Glad, 1994)

*The model $\mathcal{M}$ is s.g.i. for a given output $\boldsymbol{y}$ if every $p_i \in \boldsymbol{p}$ is s.g.i. given $\boldsymbol{y}$.*

*The model $\mathcal{M}$ is s.l.i. for a given output $\boldsymbol{y}$ if every $p_i \in \boldsymbol{p}$ is s.l.i. given $\boldsymbol{y}$.*

*The model $M$ is s.n.i. for a given output $\boldsymbol{y}$ if at least one $p_i \in \boldsymbol{p}$ is s.n.i. given $\boldsymbol{y}$.*

### 1.2.2 Model structures

The most basic model structure we consider is the SIR model. For simplicity, we assume no demography, no migration, homogenous populations and a constant, population size $N$. In this SIR model, there are three mutually exclusive compartments, each corresponding to a distinct infection state: Susceptible $S$, Infectious $I$ and Recovered (and immune) $R$. Susceptible individuals become infected at a rate $\beta I/N$ where $\beta$ is the transmission rate and is equal to the product of the contact rate and the probability that a contact will successfully result in an infection. Infectious individuals recover at a rate $\gamma$. These dynamics can be described by the following set of ODEs:

$$\textbf{SIR}: \begin{cases} \dfrac{dS}{dt} = -\dfrac{\beta SI}{N} \\ \dfrac{dI}{dt} = \dfrac{\beta SI}{N} - \gamma I \\ \dfrac{dR}{dt} = \gamma I \end{cases} \quad . \tag{2}$$

ICs for the $S$, $I$ and $R$ states will be denoted by $S(0), I(0)$ and $R(0)$, respectively. At any time $t \geq 0$, $N = S(t) + I(t) + R(t)$. For the SIR model, we consider two outputs: incidence, $y_1 = \beta SI/N$, and prevalence, $y_2 = I/N$.

For diseases with a non-negligible latent period (e.g., COVID-19 (Liu et al., 2020)), the SIR model can be modified to include a latent state $L$. The modified dynamics are described by the following set of equations:

$$\textbf{SLIR}: \begin{cases} \dfrac{dS}{dt} = -\dfrac{\beta SI}{N} \\ \dfrac{dL}{dt} = \dfrac{\beta SI}{N} - \alpha L \\ \dfrac{dI}{dt} = \alpha L - \gamma I \\ \dfrac{dR}{dt} = \gamma I \end{cases}, \quad (3)$$

where $1/\alpha$ represents the length of the latent period. Let $L(0)$ denote the IC for the latent state. For all $t \geq 0$, $N = S(t) + L(t) + I(t) + R(t)$. We study an equivalent set of outputs as for the SIR model: incidence, $y_3 = \alpha L$, and prevalence, $y_4 = I/N$.

For diseases where a relapse of symptoms is possible after a period of remission (e.g., hepatitis A), we can include a compartment $Q$ to represent the remission state. In this model, we also allow for immunity by vaccination. The dynamics of this SLIRQ (Susceptible-Latent-Infectious-Recovered (or immune)-Remission) model as adapted from Dankwa et al. (2021) are as follows. In this model, individuals in the $R$ compartment are immune, either as a result of vaccination or past infection. Susceptible individuals become exposed at a rate $\beta I/N$ and move to the latent state, where they remain for $1/\alpha$ time units, after which they become infectious. A proportion, $1 - \eta$, of infectious individuals recover temporarily, moving to the remission state for a period of $1/\sigma$ time units, after which they experience a relapse of symptoms, becoming infectious. The remaining proportion, $\eta$, of infectious individuals recover permanently and become immune. The recovery rate is $\gamma$. A number $v(t)$ of individuals are vaccinated at time $t$ and become immune. These dynamics are captured by the following set of ODEs:

**SLIRQ**:
$$\begin{cases} \dfrac{dS}{dt} = -\dfrac{\beta SI}{N} - v\dfrac{S}{N} \\[6pt] \dfrac{dL}{dt} = \dfrac{\beta SI}{N} - \alpha L \\[6pt] \dfrac{dI}{dt} = \alpha L - \gamma I + \sigma Q \\[6pt] \dfrac{dR}{dt} = v\dfrac{S}{N} + \eta\gamma I \\[6pt] \dfrac{dQ}{dt} = (1-\eta)\gamma I - \sigma Q \end{cases} \quad (4)$$

The IC corresponding to the remission state will be denoted by Q(0). For all $t \geq 0$, N = S(t) + L(t) + I(t) + R(t) + Q(t). We consider the same set of outputs as before: incidence, $y_5 = \alpha L + \sigma Q$, and prevalence, $y_6 = I/N$.

Finally, we introduce a SLIR/SLI model structure suitable for vector-borne diseases, and adapted from the works of Ngwa and Shu (2000) and Kao and Eisenberg (2018), who apply the model to malaria and dengue, respectively. In the model, infection dynamics within the host population are explained via a SLIR model, as in equation (3), while the dynamics in the vector population are explained via a SLI model, thus a SLIR/SLI model. Transmission can only occur between individuals of different populations, i.e., host-to-vector or vector-to-host. Like in the previous models, we assume constant sizes for both populations: let $N_h$ and $N_v$ represent the sizes of the host and vector populations, respectively. We use subscripts "h" and "v" to represent compartments for hosts and vectors, respectively. Thus, we have $N_h = S_h(t) + L_h(t) + I_h(t) + R_h(t)$ and $N_v = S_v(t) + L_v(t) + I_v(t)$, $\forall t \geq 0$.

The pathogen transmission rate from host to vector $\beta_{hv}$ is equal to the product of the contact rate between host and vector (in malaria for example, this may be the human biting rate of mosquitoes) and the probability of successful transmission from an infectious host to a susceptible vector. Similarly, the transmission rate from vector to host $\beta_{vh}$ is equal to the product of the contact rate between vector and host and the probability of successful transmission from an infectious vector to a susceptible host. Infected hosts become infectious after a latency period of $1/\alpha_h$ time units and remain infectious for a period of $1/\gamma_h$ time units before recovery. Recovered hosts become immune to the disease. Infectious hosts transmit the pathogen to susceptible vectors at a rate $\beta_{hv} I_h / N_h$. Infected vectors become infectious after a latency period of $1/\alpha_v$ time units. Infectious vectors transmit the pathogen to susceptible hosts at a rate $\beta_{vh} I_v / N_h$. Within each population, we assume equal birth and death rates: $\mu_h$ and $\mu_v$ for hosts and vectors, respectively, so no disease-related mortality is incorporated. The SLIR/SLI model is represented by the following system of ODEs:

**SLIR/SLI**:
$$\begin{cases} \dfrac{dS_h}{dt} = \mu_h N_h - \dfrac{\beta_{vh} S_h I_v}{N_h} - \mu_h S_h \\[4pt] \dfrac{dL_h}{dt} = \dfrac{\beta_{vh} S_h I_v}{N_h} - \alpha_h L_h - \mu_h L_h \\[4pt] \dfrac{dI_h}{dt} = \alpha_h L_h - \gamma_h I_h - \mu_h I_h \\[4pt] \dfrac{dR_h}{dt} = \gamma_h I_h - \mu_h R_h \\[4pt] \dfrac{dS_v}{dt} = \mu_v N_v - \dfrac{\beta_{hv} S_v I_h}{N_h} - \mu_v S_v \\[4pt] \dfrac{dL_v}{dt} = \dfrac{\beta_{hv} S_v I_h}{N_h} - \alpha_v L_v - \mu_v L_v \\[4pt] \dfrac{dI_v}{dt} = \alpha_v L_v - \mu_v I_v \end{cases} \qquad (5)$$

The ICs for the SLIR/SLI model will be denoted by (listed in order of states): $S_h(0), E_h(0), I_h(0), R_h(0), S_v(0), L_v(0)$ and $I_v(0)$.

The following outputs are studied: 1) incidence in hosts (host incidence), $y_7 = \alpha_h L_h$; 2) prevalence in hosts (host prevalence), $y_8 = I_h/N_h$; 3) incidence in vectors (vector incidence), $y_9 = \alpha_v L_v$; and 4) detected vector counts, $y_{10} = \lambda_v(S_v + L_v + I_v)$, $\lambda_v$ is an unknown vector detection rate.

### 1.2.3 Toolbox employed

In this study, we employ the SIAN (Structural Identifiability ANalyser) (Hong et al., 2019) software tool for structural identifiability analysis. The algorithm implemented in SIAN, proposed by Hong et al. (2020), is based on a combination of differential algebra and Taylor series approaches to structural identifiability analysis. SIAN is implemented in Maple and available as a web application: https://maple.cloud/app/6509768948056064.

Here, we are interested in assessing both local and global identifiability of model parameters, including ICs. Therefore, although other toolboxes exist which are capable of assessing the structural identifiability of $\mathcal{M}$, we employ SIAN because it uniquely possesses the following characteristics as desired for this study: 1) it is capable of assessing both local and global identifiability of model parameters; 2) it provides identifiability results for ICs of state variables; and 3) it is available as a web application and accepts a simple text-based input, hence more accessible than toolboxes which require program installation or knowledge of a particular programming language. This latter characteristic is a particularly desirable one for a structural identifiability analysis software, as it addresses a potential barrier to the application of structural identifiability analysis.

For a given model, SIAN typically produces one of the following results for the structural identifiability of each model parameter: s.g.i., s.l.i. or s.n.i. SIAN is also capable of computing identifiable combinations, although we do not employ that functionality here.

### 1.2.4 Structural identifiability assessments

Structural identifiability analysis was conducted in four stages, each stage designed to reflect a possible scenario as may be encountered when modelling infectious disease transmission. Across these stages, we studied the structural identifiability of model parameters given three common data types as model outputs – incidence, prevalence, and detected vector counts (the latter only applicable to SLIR/SLI). We analysed 26 ODE model versions, assuming in all cases constant, unknown population sizes. For each model, we assessed the structural identifiability of all unknown parameters, including ICs.

Stages are now described.

- **Stage one (single outputs, all parameters unknown)**: Structural identifiability analysis was conducted for models defined with a single data type as output and assuming all parameters were unknown. This scenario is typical in the initial stages of an outbreak of an emerging pathogen, when little is known of pathogen epidemiology and consequently, natural history parameters or transmission rates. Furthermore, in such scenarios, as data are often limited, only one type of data may be available for parameter estimation. It is therefore of interest to determine which parameters are structurally identifiable in such contexts.

  Therefore, for SIR, SLIR and SLIRQ, we assessed the structural identifiability of model parameters given either incidence or prevalence data. For SLIR/SLI, output was host incidence or host prevalence. We do not consider vector data at this stage, as such data are less likely to be available during the early stages of an emerging vector-borne disease outbreak. Thus, at this stage, eight model versions were analysed.

- **Stage two (single outputs, only natural history parameters known)**: In the case of an endemic disease which has been widely studied (e.g., malaria in sub-Saharan Africa), a high level of certainty may be obtained on the values of natural history parameters. In modelling such diseases, knowledge of natural history parameters may be assumed and hence these parameters may be treated as known quantities in the model. Stage two considers this scenario. For the model versions analysed at stage one, we assumed all natural history parameters to be known and re-evaluated the structural identifiability of the other (unknown) model parameters, i.e., all ICs, transmission rate parameters, and for the SLIR/SLI models, the birth rate parameters, additionally. This analysis enabled us to identify how the structural identifiability properties of unknown parameters change once other parameters in the model are assumed known. As in stage one, eight model versions were analysed at this stage.

- **Stage three (multiple outputs, all parameters unknown)**: In instances where surveillance capacities are strengthened in the face of an emerging outbreak, it is possible to observe more than one type of data. For example, in the context of a vector-borne disease outbreak, there may be, in addition to host incidence data, data on the size of the vector population, as could be obtained through traps in the case of mosquitoes (for mosquito-borne diseases), or field signs, in the case of badgers (for bovine tuberculosis). In stage three, we studied the structural identifiability of model parameters in these "data-rich" scenarios by defining models to have at least two output types. All parameters were treated as unknown, as in stage one. Thus, we were able to compare results obtained at this stage to results at stage one (with single outputs) to assess the influence of additional outputs on parameters' structural identifiability.

  For the SIR, SLIR and SLIRQ structures, outputs were incidence and prevalence. For the SLIR/SLI structure, we studied two output combinations. One comprised host incidence and host prevalence, reflecting a scenario in which host infection data are available but vector data are absent, while the other comprised both host and vector data: host incidence, host prevalence, vector incidence and detected vector counts. Thus, five model versions were analysed at this stage.

- **Stage four (multiple outputs, only natural history parameters known)**: Here, we consider the five model versions analysed at stage three, but assuming knowledge of natural history parameters, as in stage two. Thus, we could compare the structural identifiability of parameters at this stage to corresponding results: 1) at stage two, to determine whether additional outputs improved parameters' structural identifiability after some parameters have been assumed known; and 2) at stage three, to determine how structural identifiability of parameters improved with knowledge of natural history parameters, given multiple outputs.

## 1.3 RESULTS

Structural identifiability results of model parameters assessed at stages one, two, three, and four are presented in Table 1, Table 2, Table 3, and Table 4, respectively. For some models, SIAN was unable to complete global identifiability calculations but provides results for local identifiability. For these model versions, parameters assessed as being s.l.i. by SIAN are referred to in this study as being *at least* s.l.i., given that they may potentially be s.g.i.

Results are now discussed by stage.

**Stage one (single outputs, all parameters unknown)**: See Table 1. When all parameters were assumed unknown and single outputs considered, all models except the SLIRQ models are s.n.i. All parameters of the SLIRQ model are s.l.i., irrespective of output type. In the SIR and SLIR models with output as prevalence, the transmission rate $\beta$ is s.g.i. However, with output as incidence, $\beta$ becomes s.n.i. The IC for the recovered compartment $R(0)$ is s.n.i. in all SIR and SLIR models studied at stage one, but is at least s.l.i. in both SLIRQ models (i.e., given incidence or prevalence as output). In the SLIR/SIR model with incidence as output, the IC corresponding to the recovered compartment for hosts $R_h(0)$ is at least s.l.i. when output is host incidence but s.n.i. when output is host prevalence. The transmission rate parameter and all ICs corresponding to the vector population are s.n.i. with host prevalence or host prevalence as output, while other parameters associated with the vector population (birth rate $\mu_v$ and parameter controlling the length of latent period $\alpha_v$) are at least s.l.i.

**Stage two (single outputs, only natural history parameters known)**: See Table 2. Assuming knowledge of the natural history parameters in the SIR, SLIR and SLIR/SLI models did not lead to an improvement of the structural identifiability of parameters which were s.n.i. at stage one (where all parameters – including natural history

parameters – were unknown), irrespective of output type. However, for the SLIRQ models, the structural identifiability of unknown parameters ($\beta$ and ICs) is seen to improve with the assumption of knowledge of natural history parameters: these parameters are s.g.i. at this stage but were at least s.l.i at stage one.

**Stage three (multiple outputs, all parameters unknown)**: See Table 3. When incidence and prevalence data are considered jointly as outputs in the same model, structural identifiability of the SIR, SLIR and SLIRQ models improves considerably compared to stage one. All parameters in these models which were s.n.i. at stage one become s.g.i. For example, $\beta$ is s.n.i. in the SIR model with incidence only as output; however, with the addition of prevalence data as an output in the model, $\beta$ becomes s.g.i. Likewise, $R(0)$ is s.n.i. in all SIR and SLIR models with single outputs (either incidence or prevalence; Table 1) but becomes s.g.i. when these outputs are considered simultaneously.

For the SLIR/SLI model, all parameters associated with the host population are at least s.l.i. when host incidence and host prevalence data are joint model outputs. However, the ICs and transmission rate parameter associated with the vector population are s.n.i., as in stage one when these outputs were considered separately (Table 1).

**Stage four (multiple outputs, natural history parameters known)**: See Table 4. Even when natural history parameters are assumed known, the ICs and transmission rate parameter associated with the vector population in the SLIR/SLI model remain s.n.i. with host prevalence and host incidence as joint model outputs. It is only with the addition of vector data (vector incidence and detected vector counts) as outputs that these parameters become s.g.i.

## 1.4 Discussion

In this work, we have studied the structural identifiability of 26 ODE model versions, each with a unique combination of underlying compartmental structure (SIR, SLIR, SLIRQ or SLIR/SLI) and data type considered as model output (incidence, prevalence or detected vector counts). We gained useful insights on the influence of data types on the structural identifiability of the infectious disease transmission models studied here.

The consideration of multiple data types as outputs generally improved models' structural identifiability. Indeed, when only single outputs were considered (Table 1, Table 2), all models except the SLIRQ-structured models were s.n.i. However, when these models were defined to have at least two data types as outputs, all but one model become s.g.i. (Table 3, Table 4).

The exception – the SLIR/SLI model with outputs as host incidence and host prevalence – had its transmission rate parameter and ICs for the vector population remaining s.n.i. despite having host incidence and host prevalence as model outputs (Table 3), and even after all natural history parameters in the model were assumed known (Table 4). However, when vector-related data (vector incidence and detected vector counts) were added as outputs in the model, these parameters become s.g.i. (Table 3, Table 4), suggesting that data on host infection alone (incidence, prevalence or both) are not sufficient to identify these vector-related parameters.

We found it surprising that the other vector-related parameters studied – $\mu_v$, the vector birth rate and $\alpha_v$, the parameter controlling the length of the latent period – were at least s.l.i. given host incidence or host prevalence (Table 1), since we expected vector-related parameters to be non-identifiable in the absence of vector data. We thus checked with other structural identifiability software capable of computing global

results – GenSSI 2.0 (Ligon et al., 2018), COMBOS (Meshkat et al., 2014) and DAISY (Bellu et al., 2007) – but none of this were able to complete computations. That these vector-related parameters are identifiable with host data is not yet clear to us and it is a question we continue to explore. We suspect that these parameters are likely not practically identifiable from typically available host incidence data, even if they are structurally identifiable.

Assuming knowledge of the natural history parameters did not seem to improve the structural identifiability of parameters in the majority of single-output models (Table 1, Table 2), likely because all natural history parameters were at least s.l.i. (in those models in which they were treated as unknown parameters; Table 1), indicating that they were not in identifiable combinations. Hence, fixing the values of these parameters appeared not to have influenced existing symmetries.

We note that for all SIR and SLIR models with single outputs (incidence or prevalence), the IC corresponding to the recovered compartment $R(0)$ is s.n.i., and its structural identifiability does not improve even when natural history parameters in these models are assumed known (Table 1, Table 2). Only with the simultaneous analysis of multiple data types as outputs does $R(0)$ become s.g.i. (Table 3). It is interesting to observe this "synergy-like" effect: separately, neither incidence nor prevalence is sufficient for the identification of $R(0)$, but considered jointly, these data prove adequate to identify $R(0)$. It is of interest to study further what causes this effect in this case, as resulting insights could potentially aid to *pre-determine* which data types will lead to structural (non-) identifiability. We recommend the development of formal methods to aid such pre-determination: these methods may offer a more efficient paradigm as they may eliminate the need for trials with several data types.

Our results on the IC of the recovered state agree with Massonis et al. (2021) who, in a structural identifiability analysis of several compartmental COVID-19 transmission models, found that the recovered state is "almost never observable". That is, it is not distinguishable from other states given output and input observations, unless measured directly. An important question then arises: what sources of data are useful to inform the IC of the recovered/immune state in scenarios where this state is not directly observed? Expert knowledge or seroprevalence estimates based on representative studies may be helpful in this regard. Where these data are not readily available, this IC has often been set to zero; however, if the true value is different from zero, other parameters need to be interpreted accordingly; the transmission rate, in particular, needs to be understood as an observed transmission (i.e., the transmission estimated from the observed data) rather than a true transmission rate. The distinction may be particularly important when trying to mechanistically interpret the transmission rate as a product of constituent parameters (e.g., contact rate times probability of infection). More broadly, simulation studies and sensitivity analysis may be needed to understand the specific influences of IC values on one's parameter estimates and thus the robustness of one's inferences.

Our study is a relevant contribution to the literature as it explicitly considers ICs and population sizes as unknown in models which have been mostly studied assuming these quantities are known. Data on ICs or population size may not always be available or able to be measured directly, hence the need to study identifiability in such scenarios. Also, as we had complete control over structure-output combinations, we were able to modify model characteristics such that the cause for a change in identifiability results could be precisely identified. In addition, unlike most previous studies, we provide input code for all analyses conducted, to serve as a model to individuals who may be new to structural identifiability analysis. To further facilitate increased adoption of structural identifiability analysis, we chose to use a web-based structural identifiability analysis tool, which accepts simple text-based inputs. This eliminates potential barriers to adoption such as the need for program installation or proficiency in a programming language.

Despite these strengths, some limitations exist. First, when models were complex (i.e., having more than four states, or multiple outputs and several parameters), it was generally challenging for SIAN (and other toolboxes used) to produce complete results. More work is needed on scaling toolboxes to match the increasing complexity of modern epidemic models. Second, it would have been desirable to use multiple toolboxes for all analysis, as that would have facilitated the detection of potentially problematic results; however as stated earlier, SIAN was the only publicly available toolbox – as far as we know – which had the combination of functionalities required for this study: 1) ability to assess both local and global identifiability; 2) ability to assess identifiability of unknown ICs, and 3) possibility to implement without requiring program installation or specialized programming language skills. Work on developing more accessible toolboxes with a range of relevant functionalities is therefore warranted. Third, the selection of compartmental models studied here is limited. Similarly, although the set of data types examined here comprises some of the most commonly measured in disease surveillance, it is not representative of the wide variety of possible data types; for example, we do not consider detected incidence (i.e., incidence allowing for underreporting). Our work is intended to be primarily illustrative, providing the rationale for assessing structural identifiability and some approaches.

Our work focused on deterministic, compartmental ODE models. It would be desirable to extend our study to cover stochastic models (Browning et al., 2020); models which incorporate population structure (e.g., age-structured or spatial models); time-varying parameters, which have been shown to address structural identifiability issues due to their role in breaking symmetries in the model structure (Massonis et al., 2021a); and additional data types such as the number of recovered individuals and the number of symptomatic individuals in quarantine (for models with a quarantine compartment). A critical caveat exists, however: the available structural identifiability toolboxes only allow for deterministic ODEs, although they could be used to establish proxy identifiability results for stochastic differential equation models (Browning et al., 2020). More research is needed towards developing identifiability analysis tools suited to stochastic models.

So far, we have focused on answering the question: Given a model $\mathcal{M}$, which data types can make model parameters more structurally identifiable? Our discussions have therefore originated from an output (or data type) perspective. Less attention has been paid to the influence of the rest of the model structure on the identifiability of model parameters. The alternative question, therefore, and one that is necessary in data-limited settings, is: Which structural modifications on the system of ODEs $\dot{X}(t)$ will improve the structural identifiability of $\mathcal{M}$? Some approaches have been suggested. One approach involves reparameterizing the model with the aim to reduce the number of parameters, concentrating particularly on identifiable combinations (Eisenberg and Hayashi, 2014; Massonis et al., 2021a, 2021b; Meshkat et al., 2014; Wieland et al., 2021). Another approach centers on simplifying model complexity by reducing the number of features/states (Massonis et al., 2021a) and another entails non-dimensionalizing (Kao and Eisenberg, 2018) or scaling some state variables (Brouwer et al., 2018; Eisenberg et al., 2013). These considerations are outside the scope of the current discussion but are important to the broader goal of developing infectious disease models for useful inference.

It is important to note that although a model may be s.n.i, it may be useful for drawing inferences, if these are limited to the structurally identifiable parameters of the model (Janzén et al., 2016; Massonis et al., 2021a).

In this work, we have demonstrated the influence of data types on structural identifiability of model parameters. A careful consideration of the type of data available for parameter estimation is therefore advised as a relevant initial step in performing inference with infectious disease transmission models.

## 1.5 CONCLUSIONS

We have studied the structural identifiability of parameters of various compartmental models for infectious disease transmission. We have demonstrated the influence of data types on structural identifiability by considering different data types as model outputs and examining how structural identifiability of unknown parameters, including ICs, varied with varying outputs. The structural identifiability of some parameters varied depending on the type of model output, and single-output models were often not structurally identifiable. In general, the inclusion of additional data types as outputs improved structural identifiability of parameters. Attention ought therefore to be paid to the type(s) of observed data at hand, prior to estimating model parameters, given that data types influence a model's structural identifiability and consequently, the robustness of resulting inferences.

**Acknowledgments**

The authors thank Gleb Pogudin for helpful support on the implementation of SIAN.

**Funding**

This research did not receive any specific grant from funding agencies in the public, commercial, or not-for-profit sectors. E.A.D. was supported by a studentship at the Department of Statistics, University of Oxford. A.F.B. was supported by the National Science Foundation (grant DMS1853032) and the National Institutes of Health (grant U01GM110712). C.A.D. was supported by joint Centre funding from the UK Medical Research Council and the UK Foreign, Commonwealth & Development Office (FCDO), under the MRC/FCDO Concordat agreement and is also part of the EDCTP2 programme supported by the European Union. C.A.D. was funded on grants from the UK National Institute for Health Research (NIHR) [Vaccine Efficacy Evaluation for Priority Emerging Diseases: PR-OD-1017-20007 and HPRU in Emerging and Zoonotic Infections: NIHR200907. The views expressed in this publication are those of the authors and not necessarily those of their funding institutions.

**Table 1 (stage one)**: Structural identifiability of parameters and models assuming **all parameters are unknown** and given **single model outputs**: incidence (I) or prevalence (P). For the SLIR/SLI models, outputs corresponding to the host population are annotated with "(h)". Output cells are shaded according to the structural identifiability of the model given that output: a green shade indicates the model is structurally globally identifiable (s.g.i.), a yellow shade indicates the model is structurally locally identifiable (s.l.i.) and a brown shade indicates the model is structurally non-identifiable (s.n.i.).

| Model structure | Output | Structural identifiability of parameters | | |
|---|---|---|---|---|
| | | s.g.i. | s.l.i | s.n.i. |
| SIR | I | $\gamma, S(0), I(0)$ | | $\beta, R(0)$ |
| | P | $\beta, \gamma$ | | $S(0), I(0), R(0)$ |
| SLIR | I | $\alpha, \gamma, S(0), L(0), I(0)$ | | $\beta, R(0)$ |
| | P | $\beta$ | $\alpha, \gamma$ | $S(0), L(0), I(0), R(0)$ |
| SLIRQ | I | | $\alpha, \beta, \eta, \gamma, \sigma, S(0), L(0), I(0), R(0), Q(0)$[a] | |
| | P | | $\alpha, \beta, \eta, \gamma, \sigma, S(0), L(0), I(0), R(0), Q(0)$[a] | |
| SLIR/SLI | I (h) | | $\alpha_h, \alpha_v, \beta_h, \gamma_h, \mu_h, \mu_v, S_h(0), L_h(0), I_h(0), R_h(0)$[a] | $S_v(0), L_v(0), I_v(0), \beta_v$ |
| | P (h) | | $\alpha_h, \alpha_v, \beta_h, \gamma_h, \mu_h, \mu_v$ | $\beta_v, S_h(0), L_h(0), I_h(0), R_h(0),$ $S_v(0), L_v(0), I_v(0)$ |

[a] Parameters are at least s.l.i. No results were produced for global identifiability: SIAN timed out before global identifiability calculations could be completed.

**Table 2 (stage two)**: Structural identifiability of parameters and models assuming **all parameters except natural history parameters are unknown** and given **single model outputs**: incidence (I) or prevalence (P). For the SLIR/SLI models, outputs corresponding to the host population are annotated with "(h)". Output cells are shaded according to the structural identifiability of the model given that output: a green shade indicates the model is structurally globally identifiable (s.g.i.), a yellow shade indicates the model is structurally locally identifiable (s.l.i.) and a brown shade indicates the model is structurally non-identifiable (s.n.i.).

| Model structure | Output | Structural identifiability of parameters | | |
|---|---|---|---|---|
| | | s.g.i. | s.l.i | s.n.i. |
| SIR | I | $S(0), I(0)$ | | $\beta, R(0)$ |
| | P | $\beta$ | | $S(0), I(0), R(0)$ |
| SLIR | I | $S(0), L(0), I(0)$ | | $\beta, R(0)$ |
| | P | $\beta$ | | $S(0), L(0), I(0), R(0)$ |
| SLIRQ | I | $\beta, S(0), L(0), I(0), R(0), Q(0)$ | | |
| | P | $\beta, S(0), L(0), I(0), R(0), Q(0)$ | | |
| SLIR/SLI | I (h) | | $\beta_h, \mu_h, \mu_v, S_h(0), L_h(0), I_h(0), R_h(0)$ [a] | $\beta_v, S_v(0), L_v(0), I_v(0)$ |
| | P (h) | | $\beta_h, \mu_h, \mu_v$ [a] | $\beta_v, S_h(0), L_h(0), I_h(0), R_h(0), S_v(0), L_v(0), I_v(0)$ |

[a] Parameters are <u>at least</u> s.l.i. No results were produced for global identifiability: SIAN timed out before global identifiability calculations could be completed.

**Table 3 (stage three)**: Structural identifiability of parameters and models assuming **all parameters are unknown** and given **multiple model outputs**: outputs are incidence (I), prevalence (P) or detected vector counts (DC). For the SLIR/SLI models, outputs corresponding to the host and vector populations are annotated with "(h)" and "(v)", respectively. Output cells are shaded according to the structural identifiability of the model given that output: a green shade indicates the model is structurally globally identifiable (s.g.i.), a yellow shade indicates the model is structurally locally identifiable (s.l.i.) and a brown shade indicates the model is structurally non-identifiable (s.n.i.).

| Model structure | Output | Structural identifiability of parameters | | |
|---|---|---|---|---|
| | | s.g.i. | s.l.i | s.n.i. |
| SIR | I, P | $\beta, \gamma, S(0), I(0), R(0)$ | | |
| SLIR | I, P | $\alpha, \beta, \gamma, S(0), L(0), I(0), R(0)$ | | |
| SLIRQ | I, P | $\alpha, \beta, \eta, \gamma, \sigma, S(0), L(0), I(0), R(0), Q(0)$ | | |
| SLIR/SLI | I (h), P (h) | | $\alpha_h, \alpha_v, \beta_h, \gamma_h, \mu_h, \mu_v, S_h(0), L_h(0), I_h(0), R_h(0)$[a] | $\beta_v, S_v(0), L_v(0), I_v(0)$ |
| | I (h), P (h), I (v), DC (v) | $\alpha_h, \alpha_v, \beta_h, \beta_v, \gamma_h, \lambda_v, \mu_h, \mu_v, S_h(0), L_h(0), I_h(0), R_h(0), S_v(0), L_v(0), I_v(0)$ | | |

[a] Parameters are <u>at least</u> s.l.i. No results were produced for global identifiability: SIAN timed out before global identifiability calculations could be completed.

**Table 4 (stage four)**: Structural identifiability of parameters and models assuming **all parameters except natural history parameters are unknown** and given **multiple model outputs**: outputs are incidence (I), prevalence (P) or detected vector counts (DC). For the SLIR/SLI models, outputs corresponding to the host and vector populations are annotated with "(h)" and "(v)", respectively. Output cells are shaded according to the structural identifiability of the model given that output: a green shade indicates the model is structurally globally identifiable (s.g.i.), a yellow shade indicates the model is structurally locally identifiable (s.l.i.) and a brown shade indicates the model is structurally non-identifiable (s.n.i.).

| Model structure | Output | Structural identifiability of parameters | | |
|---|---|---|---|---|
| | | s.g.i. | s.l.i | s.n.i. |
| SIR | I, P | $\beta, S(0), I(0), R(0)$ | | |
| SLIR | I, P | $\beta, S(0), L(0), I(0), R(0)$ | | |
| SLIRQ | I, P | $\beta, S(0), L(0), I(0), R(0), Q(0)$ | | |
| SLIR/SLI | I (h), P (h) | $\beta_h, \mu_h, \mu_v, S_h(0), L_h(0), I_h(0), R_h(0)$ | | $\beta_v, S_v(0), L_v(0), I_v(0)$ |
| | I (h), P (h), I (v), DC (v) | $\beta_h, \beta_v, \lambda_v, \mu_h, \mu_v, S_h(0), L_h(0), I_h(0), R_h(0), S_v(0), L_v(0), I_v(0)$ | | |